\documentclass[final,3p,times,twocolumn]{elsarticle}
 \biboptions{comma,sort&compress}
\usepackage{here}
 \usepackage{graphicx}
  \usepackage{epsfig}

\def\nin{\noindent}
\def\beq{\begin{equation}}
\def\eeq{\end{equation}}
\def\bea{\begin{eqnarray}}
\def\eea{\end{eqnarray}}

\def\powheg{{\sc Powheg}}
\def\mcatnlo{{\sc Mc@nlo}}
\begin{document}

\begin{frontmatter}

\title{Collinearity approximations and kinematic shifts in partonic shower algorithms}

 \author[label1]{F. Hautmann}
  \address[label1]{University of Oxford, Physics Department, Oxford OX1 3NP}

 \author[label2]{H. Jung}
  \address[label2]{Deutsches Elektronen Synchrotron,  D-22603 Hamburg }

\begin{abstract}
\noindent
We  study   kinematic effects due to the approximation  of 
on-shell, collinear partons in shower Monte Carlo event generators. 
We observe that  the collinearity  approximation, combined 
  with the requirements  of  energy-momentum conservation, 
  gives rise to     a kinematic  shift, event by event,   
   in longitudinal momentum distributions.   
     We present numerical results  in the case of  jet and heavy flavor  
  production  processes measured  at the   LHC. 
\end{abstract}

\end{frontmatter}

\section{Introduction}

\nin   
Phenomenological  analyses  of complex final states produced by hard processes 
at the Large Hadron Collider (LHC) rely on event simulation by parton shower Monte Carlo 
generators~\cite{hoeche11}. These 
are used both to supplement finite-order perturbative calculations with  
 all-order,   leading-logarithmic   QCD radiative terms  
  and to incorporate 
nonperturbative effects from hadronization, multiple  parton interactions, 
underlying events~\cite{pythref,herwref}. 

The shower Monte Carlo 
generators~\cite{pythref,herwref}
treat QCD multi-parton  
 radiation within collinear ordering approximations. 
These approximations have proved to be  very  successful  
for  Monte Carlo  simulation  of    final states    at  LEP  and   Tevatron.   
On the other hand at the LHC,    unlike previous   collider experiments,       
  the phase space opens up for    events with multiple hard scales  and multiple jets to 
 occur with  sizeable  rates,   while  the  angular and rapidity  
 coverage  of detectors  extends  over a much  wider range.   
In this case   corrections to   collinear ordering  can 
  significantly   affect the structure of 
multi-jet final states~\cite{mw92,hj-ang,jhep09}.  This, 
  for instance,  will  influence   uncertainties~\cite{hoeche12}      
of  NLO-matched shower calculations~\cite{ma} for jet observables.

In this  paper    we  investigate      effects  of kinematic origin  arising from collinearity 
approximations  in the  parton showering algorithms. 
While the dynamical  corrections studied in~\cite{mw92,hj-ang,jhep09}  
 come  from 
terms   beyond NLO   in QCD perturbation theory (possibly enhanced in certain 
regions of phase space), the   
contributions  studied in this paper  are not obviously suppressed 
by powers of the strong coupling.  Rather, they 
 correspond to  approximations made  by  showering 
 algorithms  in   the parton kinematics.  They  can  be  discussed already 
  at the level 
 of leading-order and next-to-leading-order~\cite{hoeche12,ma} shower calculations.

We find  that  the collinearity  approximation, combined 
  with the requirements  of  energy-momentum conservation, 
  gives rise to     a kinematic  shift, event by event,   
   in longitudinal momentum distributions. The size of this shift  
  depends on the observable and on the 
phase space region, but becomes in general non-negligible     with increasing rapidities.  
In Sec.~2 we discuss the physical origin of this effect. In Sec.~3 we present 
 numerical illustrations  
 for jet production and $b$-flavor production based on 
   NLO event  generators.   We give  concluding remarks in Sec.~4.

\section{Longitudinal momentum shifts in parton showers}  

\nin    Consider inclusive jet  hadro-production.  LHC experiments have 
measured  one-jet cross sections~\cite{atlas-1112,cms-11-004} over a kinematic 
range in transverse momentum and rapidity  much  larger than  in 
 any previous collider  experiment.   
 Two kinds of comparisons of    
experimental data with  standard model 
theoretical predictions have been carried out.  
The first is based on NLO calculations~\cite{nagy} supplemented by 
nonperturbative  (NP) corrections estimated from shower Monte Carlo 
generators~\cite{atlas-1112,cms-11-004}.  This  shows that the NLO 
calculation agrees with data at central rapidities, while increasing deviations are 
seen with increasing rapidity at large transverse momentum $p_T$~\cite{atlas-1112}. 

A second  comparison~\cite{atlas-1112} is based on  
\powheg\  calculations~\cite{alioli}, in which 
NLO matrix elements are  matched with parton showers~\cite{herwref,pythref}.   
This data  comparison  shows    
 large differences  
in the high  rapidity region between  
results obtained by 
 interfacing 
\powheg\   with different shower  models~\cite{herwref,pythref}  and 
different model tunes~\cite{tunes}. 

The dependence of high-rapidity jet  distributions on parton showering effects 
is  discussed in~\cite{preprint}  based on    results~\cite{jhep09} 
from   high-energy factorization~\cite{hef}.   
These results are valid 
to single-logarithmic accuracy in the jet rapidity and the jet  transverse momentum, 
and  resum  terms  to  all orders  in  the strong coupling $\alpha_s$.    
In the approach~\cite{jhep09,preprint}    forward jet production is dominated by 
the scattering  of   a highly off-shell, low-$x$ parton  off a nearly on-shell, 
high-$x$ parton.  As noted in~\cite{preprint},   this  leads to 
a sizeable fraction of  reconstructed 
jets   receiving contribution from  final state partons produced by 
showering rather than just   partons   from  hard matrix elements.  

Let us    now  consider the 
NLO-matched   shower  Monte Carlo calculations   
in the light of this  physical picture.   
 The  Monte Carlo first 
generates hard subprocess events with full four-momentum assignments 
for the external lines.  In particular,  the momenta $k_j^{(0)}$  ($ j = 1 , 2 $) of the 
partons initiating the hard scatter are on shell, and are taken to be fully collinear 
with the incoming  state  momenta $p_j$, 
\begin{equation}
\label{k-before-sho} 
 k_j^{(0)} = x_j p_j  \;\;\;\;\;\;\;\;   ( j = 1 , 2 ) \;\;\; . 
\end{equation}  

  Next  the 
showering algorithm is applied, and complete final states are generated including 
additional QCD radiation from the initial-state and final-state parton cascades. 
As a result of QCD  showering, the momenta $k_j$  are no longer exactly  collinear,  
\begin{equation}
\label{k-after-sho} 
 k_j  \neq  x_j p_j  \;\;\;\;\;\;\;\;   ( j = 1 , 2 ) \;\;\; . 
\end{equation}   
This corresponds to the soft parton being far off shell in the approach~\cite{jhep09}. 
On the other hand,  in the NLO shower  calculation 
the  transverse momentum  carried by partons  with  $k_j$
is   to be   compensated  by a change  in the 
kinematics of the hard scattering subprocess. 
By  energy-momentum conservation,    this  implies  a 
reshuffling, event by event,  in  the longitudinal momentum 
fractions $x_j$ of the partons  scattering off each other in the hard  subprocess. 
The size of the shift in  $x_j$ depends on the emitted transverse momenta. 

In the next section we illustrate the longitudinal momentum 
 shifts   numerically   for jet and heavy flavor production 
  using  \powheg.

\section{Numerical results}

\begin{figure}[hbt] 
\vspace{80mm}
\includegraphics{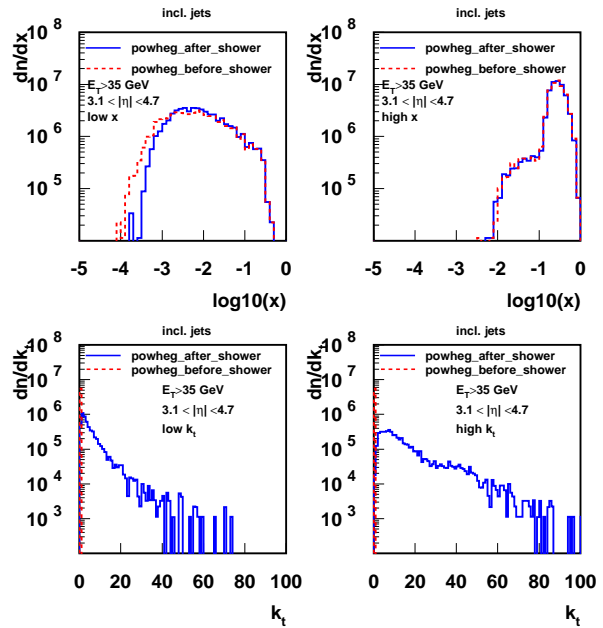}
\caption{\scriptsize Distributions in the parton longitudinal momentum fraction $x$ (upper plots) 
and  transverse momentum $k_t$  (lower plots),   before and after parton showering,  
for  inclusive jet  production  at high rapidity.}
\label{fig:fig1} 
\end{figure}

\nin   
 We   consider  
jet production in the kinematic range~\cite{atlas-1112}  and focus on the 
high rapidity region.  
In  the top panels of  Fig.~\ref{fig:fig1} we   plot the distributions  
in $x_j$  from   \powheg\  before  parton showering and after parton  showering    
  for the low-$x$ and high-$x$ partons.  
In  the bottom  panels of  Fig.~\ref{fig:fig1}   we plot the corresponding distributions in   
the transverse momentum $k_t$.

We see   from  the upper plots in  Fig.~\ref{fig:fig1}   that the  kinematical  reshuffling in the 
longitudinal momentum fraction    is negligible for  high-$x$ partons but  it is 
significant for  low-$x$ partons.    This  effect  characterizes the highly asymmetric 
parton kinematics~\cite{preprint}. 
In this  region the  kinematical shift  affects the perturbative weight 
for each event as well as the evaluation of the parton distribution functions. 
It will  be of interest to 
 examine the impact of this   phase space  
 region  on   total cross sections as well. 

The lower plots in  Fig.~\ref{fig:fig1} illustrate 
 the distribution of  transverse momenta in the initial 
state as a result of   parton showering. 
We see that this  distribution 
  falls off rapidly for the low-$k_t$  initial state parton, while  
  the distribution  for the  high-$k_t$  initial state parton  has a significant 
  large-$k_t$  tail.  This provides an  illustration,   at  the next-to-leading order,   of  
  the physical picture  described in Sec.~2  based on 
  the   single-logarithmic,  resummed  results of~\cite{jhep09,preprint}.

\begin{figure}[htbp]
\vspace{90mm}
\includegraphics{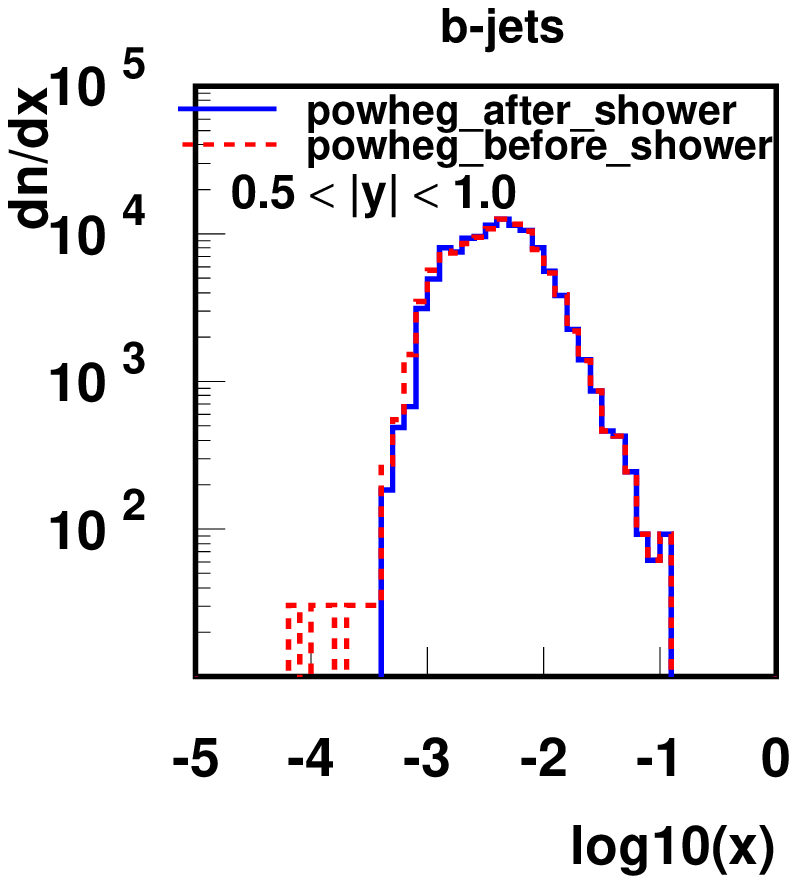}
\includegraphics{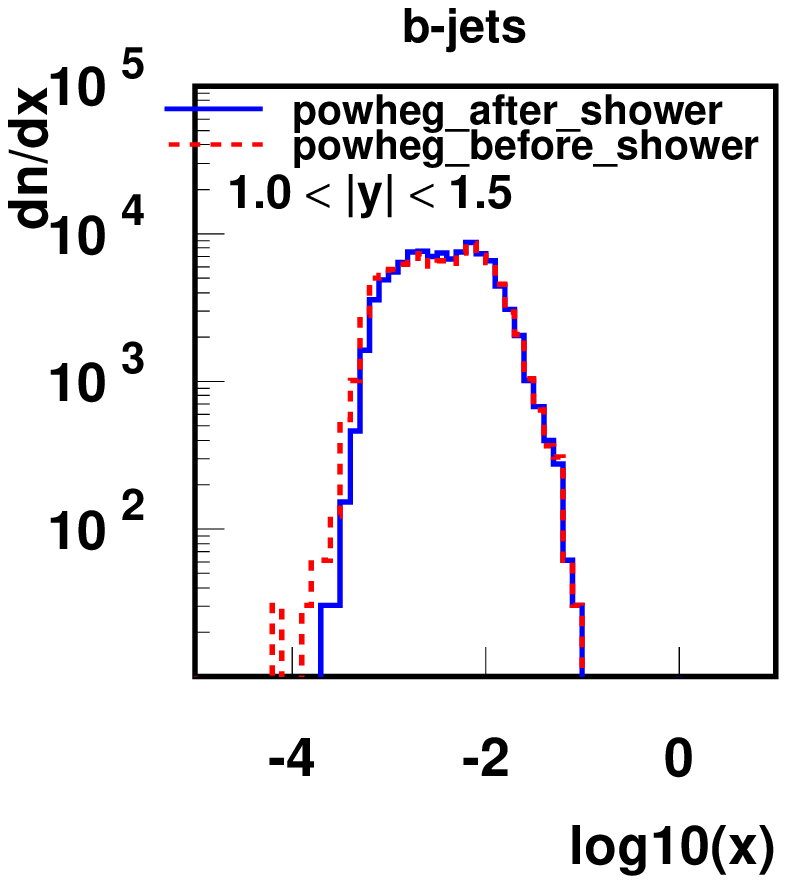}
\includegraphics{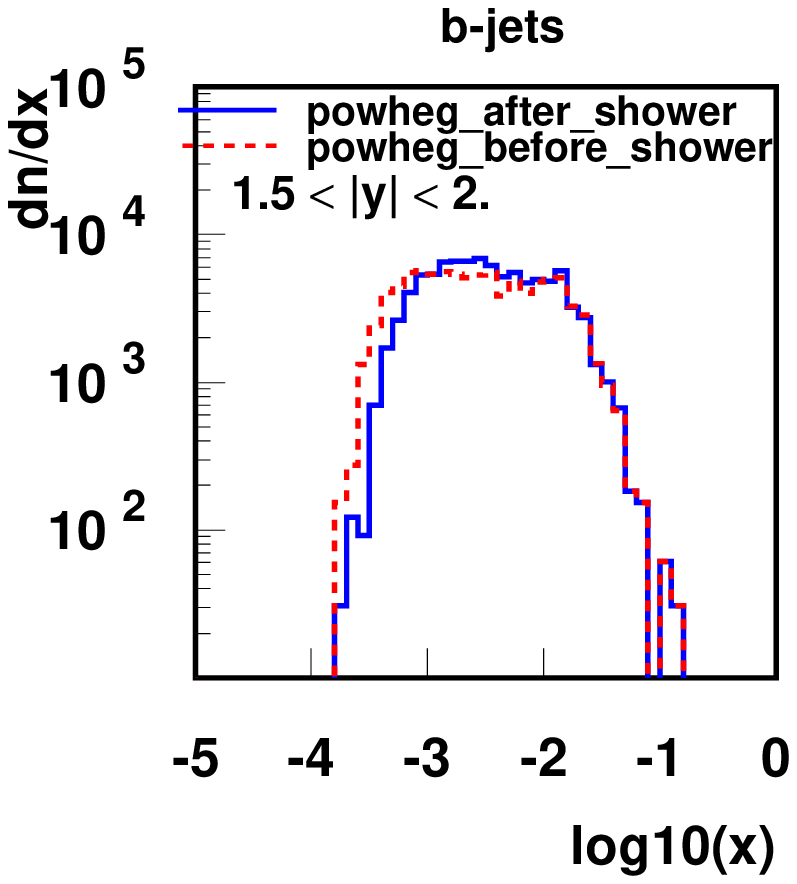}
\includegraphics{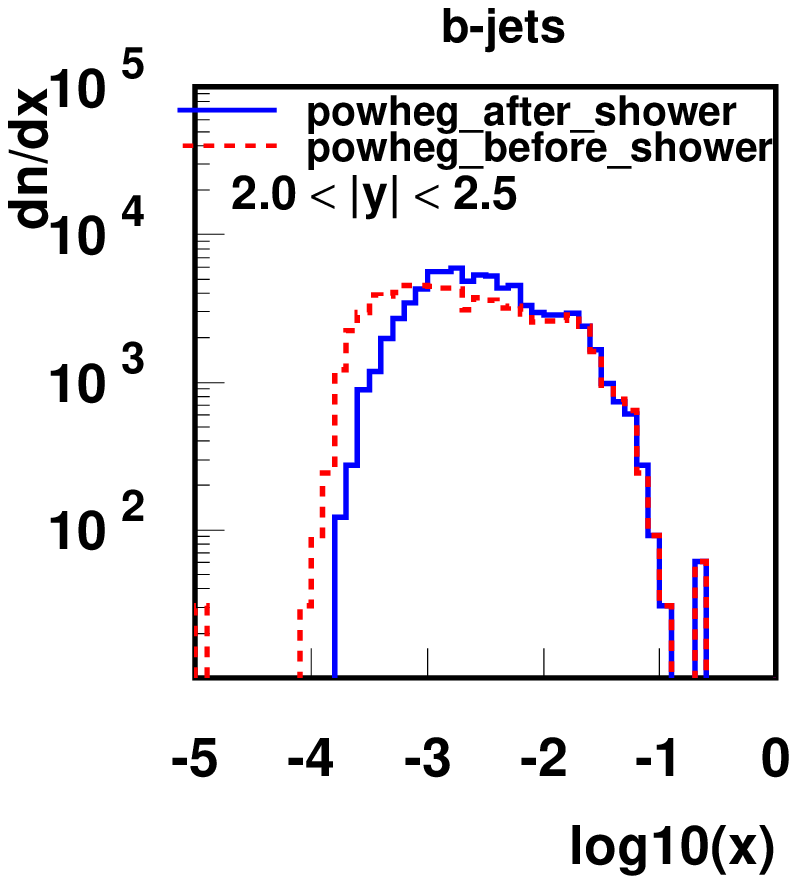}
\caption{\scriptsize  Production of $b$-jets: distribution in the parton 
longitudinal  momentum fraction $x$,  
before and after parton  showering,  for different rapidity regions.} 
\label{fig:fig2}
\end{figure}

Analogous kinematic effects  can be  examined  in 
 the case of bottom-flavor  jet  production.  
The LHC measurements of $b$-jets~\cite{cms-bjet,atlas-bjet} 
 are reasonably described by NLO-matched 
shower generators  \mcatnlo~\cite{frix}  and \powheg~\cite{frix-pow}  
 at central rapidities, and 
 they are below these predictions at large rapidity and large $p_T$. 
In  Fig.~\ref{fig:fig2}  we   consider 
  $b$-jets   for several  different rapidity intervals~\cite{cms-bjet}  and 
plot the  gluon $x$  distribution  
 from   \powheg\  before  parton showering and after parton  showering. 
 We observe similar shift    in  longitudinal 
 momentum 
  with increasing rapidity  as in the inclusive  jet case.

The  longitudinal momentum shifts  
computed in this  section 
may   affect    parton distribution functions in 
parton shower calculations. 
Although we have illustrated the shifts using the event generator 
\powheg,  the effect is   common 
 to  other   calculation methods  using  collinear shower  algorithms. 
Further studies  of this effect   will be reported elsewhere.

\section{Conclusions}   

\nin   In this paper  we have  observed that,    in   particular 
phase space regions  for  which    inclusive  jets  and $b$-jets are measured 
at the LHC,       QCD  parton showers     
are sensitive  to kinematic corrections associated with 
approximations of on-shellness and  collinearity  on  the  partonic   states 
 to which branching algorithms   are applied.

Kinematic reshuffling  in the longitudinal 
momentum fractions comes from the implementation of energy-momentum 
conservation in collinearly-ordered  showering algorithms.   
It depends on the emitted transverse momenta, and   is 
present at leading order as well as at next-to-leading order. 
We have illustrated  this  effect  quantitatively 
 using   NLO-matched event generators. 

  While the  effect is 
generally small for central production if the center-of-mass    energy is  
not too high, 
we have found that it becomes sizeable  in  kinematic  regions 
probed by  hard  production processes   at the LHC, for instance jet and 
heavy-flavor production  at high rapidity.   
These regions have not been  accessed experimentally  before  at  previous colliders  
 (e.g., at the Tevatron). 

The event-by-event  shift in longitudinal momentum distributions 
 affects  both perturbative weights and parton distribution functions. It  
contributes  in particular to the  uncertainty   on   predictions  which incorporate 
 results of    collinearly-ordered  showering algorithms.

Formulations  that keep track of non-collinear  (i.e., transverse and/or 
anti-collinear)  momentum components 
using unintegrated parton 
distributions~\cite{muld-rog-rev,unint09,avsar11}      
will be helpful  to  take account of  the  kinematic effect of longitudinal momentum shifts.   
It will also 
be of interest to investigate the relative size of the kinematic contribution 
studied in this article  with respect to 
  high-energy dynamical~\cite{mw92,hj-ang,jhep09}  effects 
 on jet final states which  can be  included   by  unintegrated formalisms.

%%%%%%%%%%%%%%%

\end{document}